\documentclass[5p]{elsarticle}

\usepackage{lineno,hyperref}
\modulolinenumbers[5]
\usepackage{subcaption}
\usepackage{amsmath}
\usepackage{caption}
\usepackage{graphicx}
\usepackage{booktabs}
\usepackage{comment}

\newcommand{\f}{\frac}
\newcommand{\p}{\partial}

\renewcommand{\v}[1]{\mathbf #1}

\journal{Computer Physics Communications}

\begin{document}

\begin{frontmatter}

\title{SPACE: 3D Parallel Solvers for Vlasov-Maxwell and Vlasov-Poisson Equations for Relativistic Plasmas with Atomic Transformations}

\author[BNL]{Kwangmin Yu}
\author[LBL]{Prabhat Kumar}
\author[SBU]{Shaohua Yuan}
\author[SBU]{Aiqi Cheng}
\author[SBU,BNL]{Roman Samulyak\corref{Roman Samulyak}}
\cortext[Roman Samulyak]{Corresponding author}
\ead{roman.samulyak@stonybrook.edu}

\address[BNL]{Computational Science Initiative, Brookhaven National Laboratory, Upton, New York 11973, USA.}
\address[LBL]{Center for Computational Sciences and Engineering, Lawrence Berkeley National Laboratory, Berkeley, CA 94720 , USA.}
\address[SBU]{Department of Applied Mathematics and Statistics, Stony Brook University, Stony Brook, New York 11794, USA.}

\begin{abstract}
A parallel, relativistic, three-dimensional particle-in-cell code SPACE has been developed for the simulation of electromagnetic fields, relativistic particle beams, and plasmas. In addition to the standard second-order Particle-in-Cell (PIC) algorithm, SPACE includes efficient novel algorithms to resolve atomic physics processes such as multi-level ionization of plasma atoms, recombination, and electron attachment to dopants in dense neutral gases. SPACE also contains a highly adaptive particle-based method, called Adaptive Particle-in-Cloud (AP-Cloud), for solving the Vlasov-Poisson problems. It eliminates the traditional Cartesian mesh of PIC and replaces it with an adaptive octree data structure. 
The code's algorithms, structure, capabilities, parallelization strategy and performances have been discussed. Typical examples of SPACE applications to accelerator science and engineering problems are also presented.
\end{abstract}


\end{frontmatter}

\tableofcontents

\section{Introduction}
The study of high intensity lasers or relativistic particle beams interacting with plasmas is important for numerous applications ranging from high energy density physics and accelerator science and technology to industrial applications. With the advent of new laser technologies, there has been a dramatic increase in achieved laser energies and intensities over the last decade. Electromagnetic fields of such high intensities lead to a variety of complex, highly nonlinear, and relativistic processes in plasma \cite{Umstadter2003}. Similarly, relativistic particle beams traveling in neutral gases induce the ionization of gas molecules and the formation of plasma, a spectrum of atomic processes, and wakefields in plasma \cite{Esarey2009,Albert2021}. 
An accurate description of such processes and understanding of the interplay between them in a 
self-consistent manner requires high-fidelity computational methods and tools.

Particle-in-Cell methods have been an attractive choice in the field of computational plasma physics due to their simplicity and ability to self-consistently resolve highly nonlinear physics processes. Over last few decades a number of PIC codes have been developed, tested, and applied to the simulation of processes in plasmas. These include OSIRIS \cite{Hemker2002}, VORPAL \cite{Nieter2004}, Warp-X \cite{Vay2018}, and QuickPIC \cite{Huang2006} to name a few. While the field of numerical simulations of plasmas interacting with lasers and relativistic particle beams is relatively mature, new quantitative and qualitative algorithms are needed to address challenges of real world applications. These are associated with the need to resolve new physics processes, achieve sufficient numerical resolution in large 3D domains, perform simulations over long time intervals, and take full advantage of modern supercomputer architectures. 

 In this paper, we describe a parallel, fully relativistic, three-dimensional particle-in-cell code SPACE for the simulation of electromagnetic fields, relativistic particle beams, and plasmas with the focus on atomic physics processes. In addition to the standard PIC algorithm, SPACE includes new atomic physics algorithms, an efficient method for highly relativistic beams in nonrelativistic plasma, support for simulations in relativistic moving frames, and a special data transfer algorithm for transformations from moving frames to laboratory frames that resolves the problem of individual times of particles. SPACE contains libraries of atomic physics processes such as the generation and evolution of plasma, including multiple ionization of ions, recombination of plasma, and electron attachment on dopants in dense neutral gases. The code also contains a highly adaptive and artifact-free particle method, called Adaptive Particle-in-Cloud (AP-Cloud) \cite{Wang2016}, for solving Vlasov-Poisson problems with optimal numerical resolution.  

Since its development, SPACE has been successfully used in several projects in the area of plasma science and particle accelerator science and engineering. These include experiments on high-pressure radio frequency cavity at Fermi National Accelerator Laboratory \cite{Yu2017,Yu2018}, coherent electron cooling program at Brookhaven National Laboratory \cite{Ma2018}, and the laser wakefield acceleration experiments driven by CO\textsubscript{2} laser at the Accelerator Test Facility (ATF) of Brookhaven National Laboratory \cite{Kumar2019,Kumar2021}. The AP-Cloud module of SPACE has been extensively used for the simulation of coherent electron cooling of relativistic ion beams \cite{Ma2018,Ma2019}.

The remainder of the paper is organized as follows. In section \ref{Models}, mathematical models and numerical algorithms implemented in the code will be described in detail. In section \ref{APCloud}, main ideas of  the Adaptive Particle-in-Cloud algorithm for solving Vlasov-Poisson problems with automatically adjusted optimal numerical resolution is presented. Atomic physics algorithms such as ionization by high energy particle beams and lasers, recombination of plasma, and electron attachment on dopant molecules will be presented in section \ref{atomic}.  Implementation of various algorithms, code structure and parallelization strategies for the use of multi-core machines will be detailed in section \ref{Implement}. Some typical applications of SPACE are presented in section \ref{Apply}. We conclude the paper with a summary of our results and perspectives for the future work.

\section{\label{Models}Models and algorithms}

\subsection{\label{Equations}Main Governing Equations} 

The standard governing equations for collisionless plasma are the Vlasov-Maxwell equations \cite{Henon_82} for distribution functions 
$f_e(\v r, \v p, t)$ and $f_i(\v r, \v p, t)$ for electrons and (positive) plasma ions:
\begin{align}
&\f {\p f_e}{\p t} + \v v_e \cdot \nabla_{\v r} f_e - e \left( \v E + \v v_e \times \v B\right)\cdot \nabla_{\v p} f_e =0, \\
&\f {\p f_i}{\p t} + \v v_i \cdot \nabla_{\v r} f_i + Z_i e \left( \v E + \v v_i \times \v B\right)\cdot \nabla_{\v p} f_i =0, \\
&\f{\p \v E}{\p t} - \f1{ \epsilon_0 \mu_0} \nabla_{\v r}\times\v B= -\f1{\epsilon_0}\v J,\label{eq:Et}\\
&  \f{\p \v B}{\p t} + \nabla_{\v r}\times\v E= 0,\label{eq:Bt}\\
&\nabla_{\v r}\cdot\v E= \f1{\epsilon_0} \rho,\\
&\nabla_{\v r}\cdot\v B= 0.
\end{align}
Here $\v r$ and $\v p$ are vectors of coordinates and momenta, $e$ is the electron charge, $Z_i$ is the ionization state of an ion, 
and $\v E$ and $\v B$ represent collective self-consistent electromagnetic field created at point $\v r$  at time moment $t$ by all plasma particles. The charge density, electric current density, and particle velocity are computed as follows:
\begin{align*}
&\rho = e\int \left( Z_if_i - f_e\right) d^3 p, \\
&\v J = e\int \left( Z_if_i \v v_i- f_e \v v_e\right) d^3 p, \\
&\v v_{\alpha} = \f {\v p/m_{\alpha}}  { \sqrt{ 1+ (p/m_{\alpha} c)^2}},
\end{align*}
where $\alpha$ denotes charged particle species.
 
In the case of non-relativistic, zero-magnetic field limit, the  Vlasov-Maxwell equations reduce to the Vlasov-Poisson system of equations

\begin{align}
&\f {\p f_{\alpha}}{\p t} + \v v_{\alpha} \cdot \nabla_{\v r} f_{\alpha}+ Z_{\alpha} e \left( \v E + \v v_e \times \v B\right)\cdot \nabla_{\v p} f_{\alpha} =0, \label{eq:VP1} \\
&\nabla_{\v r}^2\phi = -\f1{\epsilon_0} \rho,
\label{eq:VP2}
\end{align}
where $\phi$ is the electrostatic potential, 
$\v E = -\nabla_{\v r}\phi$, $Z_{\alpha}$ is the charge (in terms of the elementary charge) for particle species, and $Z_{\alpha} =-1$ for electrons.

A standard numerical method for solving the Vlasov-Maxwell and Vlasov-Poisson system of PDE's is the particle-in-cell (PIC) method \cite{HockneyEastwood}. In PIC, a fixed mesh is used for the discretization of electromagnetic fields, and the particle density distribution function is represented by Lagrangian particles that move through the mesh by the Newton-Lorentz force equation,  generating electric current. Since the PIC method eliminated the need to compute derivatives in the momentum space, we omit the index $\b r$ at the operator $\nabla$ and assume in the reminder of this paper that it represents derivatives in the geometric space.  

The core  PIC algorithm for solving the Vlasov-Maxwell equations in SPACE is typical and similar to most of PIC codes; for completeness, some details of its implementation in SPACE are presented in the next section. 
In addition to the standard PIC implementation of the Vlasov-Poisson system of PDE's, SPACE implements a novel Adaptive Particle-in-Cloud (AP-Cloud) method \cite{Wang2016}, described in Section \ref{section:apcloud}.
 
\subsection{Finite Difference Time Domain Method}
\begin{figure}[h!]
	
	\begin{subfigure}{0.49\linewidth}
		\centering
		\includegraphics[width=\linewidth]{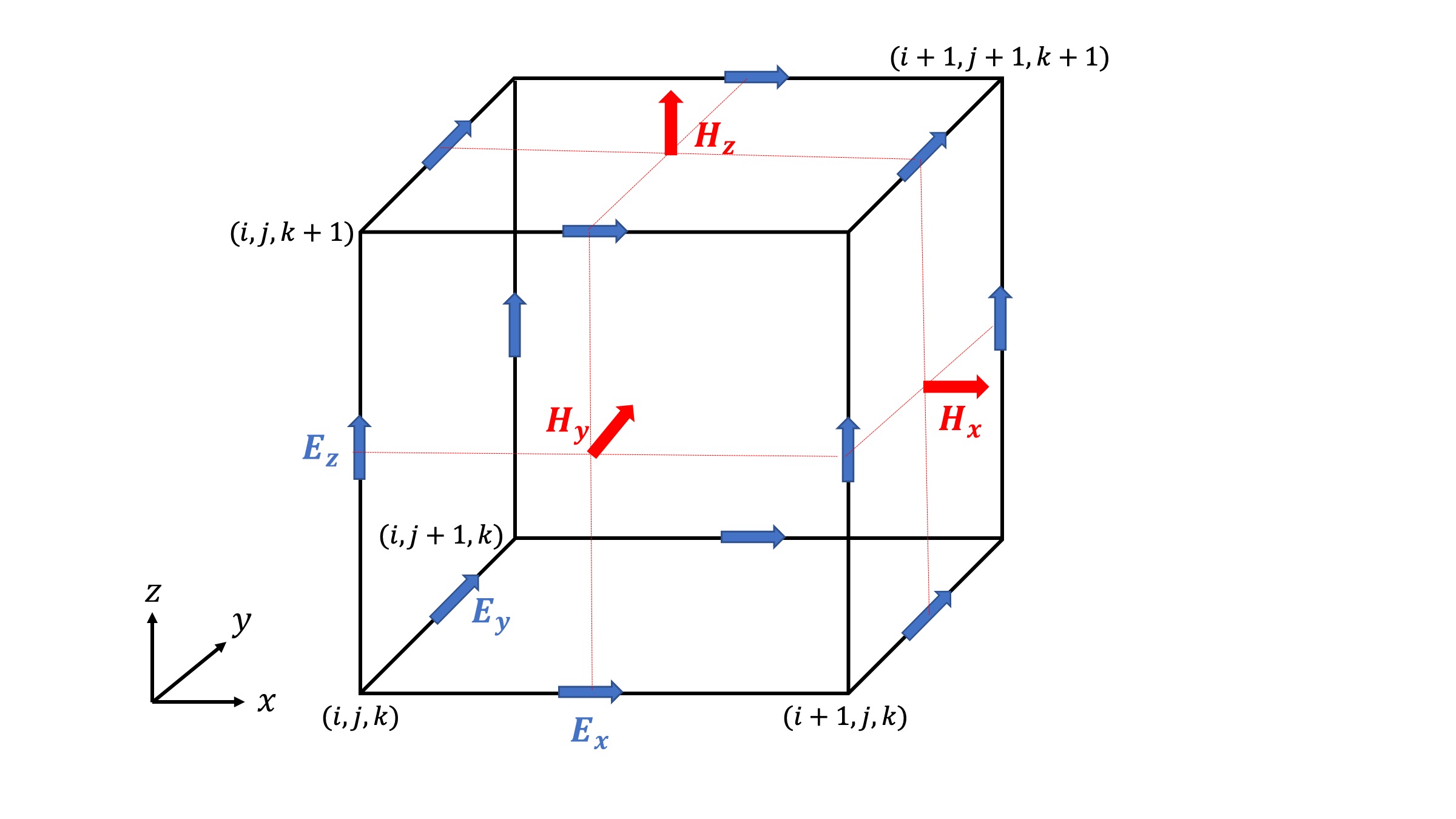}
		\caption{}
		\label{yee_cell}
	\end{subfigure} 
	\begin{subfigure}{0.49\linewidth}
		\centering
		\includegraphics[width=\linewidth]{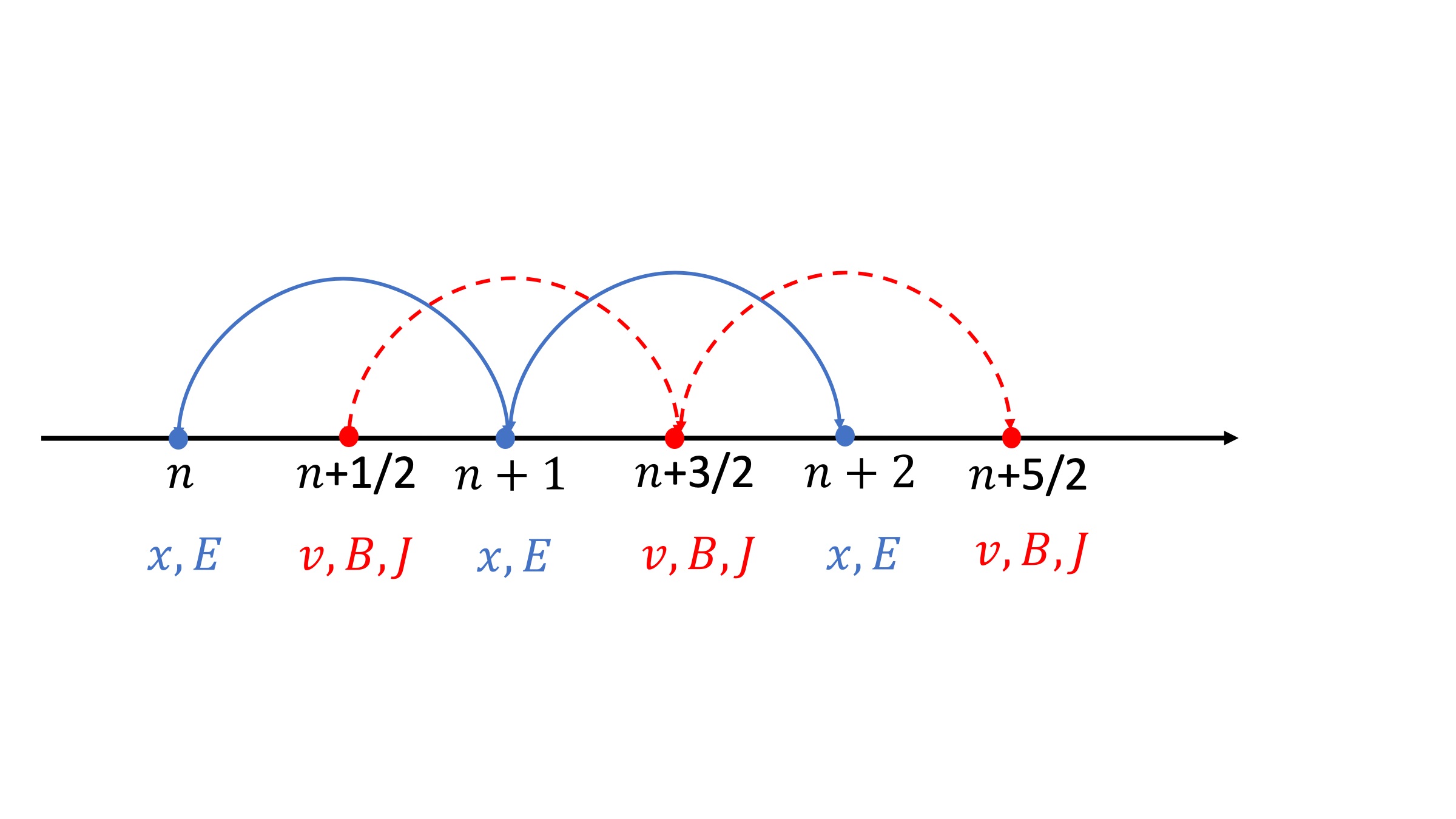}
		\caption{}
		\label{cycle}
	\end{subfigure} 
	
	\caption{(a) Yee cell and the location of EM field components as implemented in SPACE and (b) Computation sequence in time.}
	\label{Yee cell}
\end{figure}

The finite difference time domain (FDTD) method \cite{Yee1966} has been a common choice for solving Maxwell's equations numerically. It employs the Yee grid, where the electric field components are located on the edges and the magnetic field components on the faces of the cell. The location of components of the EM field and the computation sequence in time, as implemented in SPACE, are shown in figure \ref{Yee cell}.

Equations (\ref{eq:Et}) and (\ref{eq:Bt}) are discretized using second order accurate central difference approximations and the resulting finite difference equations are advanced in a leapfrog manner:
\begin{equation}
\frac{\mathbf{E}^{n+1} - \mathbf{E}^{n}}{\Delta t} =  \frac{1}{\mu_0\epsilon_0}\nabla\times\mathbf{B}^{n+1/2} - \frac{1}{\epsilon_0}\mathbf{J}^{n+1/2}
\end{equation}

\begin{equation}
\frac{\mathbf{B}^{n+1/2} - \mathbf{B}^{n-1/2}}{\Delta t} = -\nabla\times\mathbf{E}^{n+1/2}.
\end{equation}
Here different components of $\mathbf{E}$, $\mathbf{B}$ and $\mathbf{J}$ are defined at different grid points and at different times as

\begin{equation}
\mathbf{E}^n = (E_{i+1/2,j,k}, E_{i,j+1/2,k}, E_{i,j,k+1/2})^n
\end{equation}

\begin{equation}
\mathbf{B}^{n+1/2} = (B_{i,j+1/2,k+1/2}, B_{i+1/2,j,k+1/2}, B_{i+1/2,j+1/2,k})^{n+1/2}
\end{equation}
and 
\begin{equation}
\mathbf{J}^{n+1/2} = (J_{i+1/2,j,k}, J_{i,j+1/2,k}, J_{i,j,k+1/2})^{n+1/2}.
\end{equation}
where $(i,j,k)$ are grid indices for $(x,y,z)$ coordinates, respectively, and the time step size $\Delta t$ satisfies the CFL stability condition $c\Delta t < \sqrt{\Delta x^2 + \Delta y^2 + \Delta z^2}$.

\subsection{Dynamics of Particles and Current}

The particle mover solves the Newton-Lorentz equation to advance each macroparticle using the EM field calculated by the field solver:
\begin{equation}
\frac{d\mathbf{p}_\alpha}{dt} = q_\alpha(\mathbf{E}(\mathbf{x}_\alpha) + \mathbf{v}_\alpha\times\mathbf{B}(\mathbf{x}_\alpha)).
\label{NewtLor}
\end{equation}
Here $\mathbf{p}_\alpha = \gamma m_\alpha v_\alpha$ is the relativistic momentum of a particle with mass $m_\alpha$,
velocity $v_\alpha$, and charge  $q_\alpha$ at the location $\mathbf{x}_\alpha$, where $\gamma = 1/\sqrt{1-v_\alpha^2/c^2}$ is the relativistic factor.

EM field values at the location of each particle are obtained by interpolating the field components from the grid to the particle location using a second order interpolation scheme to maintain the second order overall accuracy. Equation \ref{NewtLor} is solved numerically using the following leapfrog discretization : 
\begin{equation}
\frac{\mathbf{x}^{n+1}_\alpha - \mathbf{x}^n_\alpha}{\Delta t} = \mathbf{v}^{n+1/2}_\alpha
\label{leapfrog1}
\end{equation} 

\[
\frac{\mathbf{p}^{n+3/2}_\alpha - \mathbf{p}^{n+1/2}_\alpha}{\Delta t} = q_\alpha(\mathbf{E}^{n+1}(\mathbf{x}^{n+1}_\alpha) + 
\]
\begin{equation}
\frac{\mathbf{v}^{n+3/2}_\alpha + \mathbf{v}^{n+1/2}_\alpha}{2}\times\mathbf{B}^{n+1}(\mathbf{x}^{n+1}_\alpha))
\label{leapfrog2}
\end{equation} 
This algorithm has been implemented following the correction to the Boris scheme \cite{Boris_scheme} in the method proposed by J.-L. Vay \cite{Vay}. Vay's particle pusher is $O(\Delta t^2)$ and provides a Lorentz invariant velocity update.

{\bf Current Deposition}

Since each macroparticle in a PIC method typically represents a large number of real particles, it is necessary to choose a spatial distribution of particle weighting throughout the volume occupied by a macroparticle. PIC codes employ shape functions
belonging to a family of B-spline basis functions. In SPACE, one of the three lowest order shape functions can be selected : nearest grid point (NGP), cloud-in-cell (CIC), and triangle shaped cloud (TSC). The TSC shape function defined in equation (\ref{TSC}) maintains the overall second order accuracy.
\begin{equation}
W^2(x_i) = \begin{cases}
\frac{3}{4} - \delta^2 & \text{if } |\delta| < \frac{1}{2}, \\
\frac{1}{2}(\frac{3}{2} - |\delta|)^2 & \text{if } \frac{1}{2}\leq |\delta| < \frac{3}{2}, \\
0 & \text{otherwise},
\end{cases}
\label{TSC}
\end{equation}
where $x_i$ is the location of a grid point, $x_p$ is the location of a macroparticle, and $\delta = (x_p - x_i)/\Delta x$.

In order to maintain the momentum conservation and avoid artificial self-forces, the computation of field values at particle positions and the charge density on the mesh must use the same interpolation technique. In the second order method, the charge density on the mesh is calculated using the following expression 
\begin{equation}
\rho(x_i) = \sum_{p}q_pn_pW^2(\mathbf{x}_p - x_i)
\label{rhoW}
\end{equation}
where $q_p$ is the charge of the macroparticle and $n_p$ its number density.

In SPACE, currents produced by the motion of charged macroparticles are calculated using two different methods. For CIC shape functions, the code implements the rigorous charge conservation algorithm developed by Villasenor and Buneman \cite{VilBun}. Higher order shape functions are known to reduce discrete particle noise significantly. Density decomposition method proposed by Esirkepov \cite{Esirkepov}, which is a generalization of the method developed by Villasenor and Buneman \cite{VilBun}, works with arbitrary shape functions of any order. Currently SPACE implements Esirkepov's method for second order shape functions and can be extended to support higher order shape functions. Current density thus calculated, satisfies the discrete form of the continuity equation $\frac{\partial \rho}{\partial t} +  \nabla\cdot\mathbf{J} = 0$. As a consequence, the divergence of $\mathbf{E}$ always remains equal to $\rho/\epsilon_0$ and the divergence of $\mathbf{B}$ remains zero.

\subsection{\label{APCloud}AP-Cloud method}
\label{section:apcloud}

In this section, we describe numerical algorithms for solving the Vlasov-Poisson system (\ref{eq:VP1} - \ref{eq:VP2}). In addition to  the traditional electrostatic PIC method, widely used in the majority of codes, SPACE implements a novel Adaptive Particle-in-Cloud (AP-Cloud) method \cite{Wang2016} for optimal solutions of the Vlasov-Poisson equations.  AP-Cloud is an adaptive, fully particle-based replacement of PIC. AP-Cloud replaces the Cartesian grid in the traditional PIC with adaptive computational nodes or particles, to which charges of physical macroparticles are assigned by a weighted least-square approximation. The partial differential equation is then discretized using a generalized finite difference (GFD) method and solved with fast linear solvers. The density of computational particles is chosen adaptively, so that the error of GFD and that of the source integration are balanced and the total error is approximately minimized. The method is independent of geometrical shape of computational domains and free of artificial parameters. The development of AP-Cloud was motivated by the problem of achieving an optimal adaptive resolution for numerical solutions of the Vlason-Poisson system containing large variations of sources (charges).

\subsubsection{Error Analysis of Traditional PIC Method}

The convergence of traditional PIC depends on the balance of the Laplacian operator discretization error, which reduces with the mesh refinement, and the Monte-Callo error of the charge evaluation, which increases with the mesh refinement if the total number of macroparticles remains constant. The total error for the second-order finite difference discretization is
\begin{equation}
O\left( (N\Delta x^D)^{-1/2} \right) + O(\Delta x^2),
\end{equation}
where $N$ is the number of macroparticles, $\Delta x$ is the cell size, and $D$ is the space dimension. The error is minimized \cite{Wang2016}  if
\begin{equation}
\Delta x = O\left( \left(\rho(x_k) N\right)^{-\frac1{4+D}} \right),
\label{error_PIC}
\end{equation}
where $\rho$ is the charge density at grid point $x_k$.
Since it is impossible to achieve such a balance if a uniform mesh is used for a highly nonuniform particle distribution, the use of block-structured adaptive mesh refinement approaches (AMR-PIC) has been investigated \cite{Vay2002,Vay2004}. It has been shown  \cite{Vay2002} that AMR-PIC suffers from significant numerical artifacts that exhibit themselves in unphysical image charges reflected by physical particles across interfaces between coarse and fine mesh patches. A mitigation method for this phenomenon was proposed in  \cite{Colella2010}. 
AP-Cloud method achieves an optimal numerical resolution without suffering from artifacts described in \cite{Vay2002}. The analog of formula (\ref{error_PIC}) for AP-Cloud is given in the next section.

\subsubsection{Main algorithms of AP-Cloud}

AP-Cloud uses computational particles (nodes) instead of Cartesian grid, the distribution of which is derived using an error balance criterion. Instead of the finite difference discretization of the Laplace operator, the framework of weighted least squares approximation, also known as the generalized finite-difference (GFD) \cite{BenitoUrena2001}, is applied. The framework includes interpolation, least squares approximation, and numerical differentiation on a stencil in the form of cloud of computational nodes in a neighborhood of the point of interest. It is used for the charge assignment scheme, numerical differentiation, and interpolation of solutions. 
  
 The Particle-in-Cloud method operates as follows:
\begin{itemize}
\item Given a distribution of physical macro-particles, optimally select a subset of computational nodes from this distribution by constructing an octree and applying the error balance criterion
\begin{equation} \label{errorbalance}
h = O\left( \left(\rho(x_k) N\right)^{\frac1{2(1-P)-D}} \right),
\end{equation}
where $h$ is the local averaged distance between computational nodes and $P$ is the order of interpolation polynomial in the GFD method.
\item Place computational particles on the boundary 
\item Enforce the 2:1 balance of inter-node distances in the case of extreme density changes. The 2:1 balance requires that the difference between the levels of refinement of two neighbors is at most one, improving the smoothness in the placement of computational particles. 
\item Assign physical states to computational nodes and approximate differential operators in the location of computational nodes using GFD.
\item Solve the corresponding linear system using a fast parallel solver.
\item Calculate the solution gradient on computational nodes using the same GFD stencils.
\item Interpolate gradients back to the location of macroparticles using Taylor expansion.
\end{itemize}

\subsection{\label{atomic}Atomic physics algorithms}

A distinct feature of SPACE is its ability to resolve the dynamics of complex atomic transformations, sometimes called plasma chemistry. SPACE algorithms support ionization and recombination, multi-level ionization in high-Z gases, secondary ionization by electron impact, electron attachment to neutral atoms with the formation of negatively changed ions and the subsequent ion-ion recombination etc. Three examples of distinct atomic physics processes are given below.

 \subsubsection{Long-time scale atomic processes in gaseous mixtures}
 
 Consider, for example, a high-energy proton beam propagating through dense hydrogen gas that contains a small amount of oxygen as dopant. This example is relevant to the representative SPACE application to the plasma loading of high-pressure RF cavities, described in more detail in Section \ref{Apply}. 
When an electronegative gas such as oxygen is added to the hydrogen  gas, a three-body attachment process takes place in the plasma, which is significantly
faster than the electron - ion recombination \cite{Freemire_thesis}.
The negative ions produced by the attachment process recombine with hydrogen ions. The governing equations are
\begin{align}
\notag \frac{d n_e}{d t} &= \dot{N_e} - \beta_e {n_e} n_{H^+} - \frac{n_e}{\tau}\\
\frac{d n_{H^+}}{d t} &= \dot{N_e} - \beta_e n_e n_{H^+} - \eta n_{H^+} n_{O_2^-}\\
\notag \frac{d n_{O_2^-}}{d t} &= \frac{n_e}{\tau} - \eta n_{H^+} n_{O_2^-}
\end{align}
where $n_e$, $n_{H^+}$, and $ n_{O_2^-}$ denote the number density of electrons, positive hydrogen ions, and negative dopant ions, respectively, 
and  $\dot{N_e}$ denotes the production rate of electrons. $\beta_e$ is the recombination rate of electrons on ions, $\tau$, $\eta$, and $n_{O_2^-}$ are  the attachment time, effective ion - ion recombination rate, and the number density of dopant gas ions, respectively.
The averaged hydrogen ion cluster that represents the sum  $\sum_j \beta_j n_{{H_j}^+}$ is denoted as $\beta_e n_{H^+}$.

As the recombination and attachment rates depend on  the external field, they are functions of spatial coordinates and time.
The attachment time and the ion - ion recombination rate have been measured experimentally, but only over a narrow range of RF field amplitudes.
Based on measured values, simulations establish accurate fit functions for the attachment time and the ion - ion recombination rate.

\subsubsection{Laser tunneling ionization in gases}

Laser field ionization effects such as the creation of plasma through tunneling ionization, frequency up-shifting of the laser, harmonic generation and scattering instabilities have been shown to have important macroscopic effects in laser-matter interaction studies\cite{doi:10.1063/1.1566027}. Several ionization models have been proposed and implemented in PIC codes and comparative studies have been performed and it has been shown that different models are more accurate in different cases \cite{CHEN2013}. 

Tunneling ionization algorithms based on the ADK formula have been implemented in SPACE. 
For the general case of multi-level ionization of high-atomic-number gases, the ionization probability is given by the following ADK formula \cite{Ammosov1986,CHEN2013}
\begin{eqnarray}\label{eq:multi-level-W}
	W_{lm} =&&\omega_\alpha \sqrt{\frac{3{n^*}^3 E_L}{\pi Z^3 E_a}} \frac{Z^2}{2 {n^*}^2} \left(\frac{2e}{n^*}\right)^{2n^*} \nonumber \\
	&&\exp{\left[-\frac{2E_a}{3E_L} \left(\frac{Z}{n^*}\right)^3\right]}\times
	\\		
	&&\frac{(2l+1)(l+|m|)!}{2\pi n^* 2^{|m|} (|m|)!(l-|m|)!} \left(2\frac{E_a Z^3}{E_L {n^*}^3}\right)^{2 n^*-|m|-1} \nonumber
\end{eqnarray}
Here, $\omega_a = \alpha^3 c/r_e =4.13\times10^16 s^{-1}$ is the atomic unit frequency. $E_a = 510 GV/m$ is the atomic unit of Electric field. $E_L\ (GV/m)$ represents the local strength of laser pulse. $l$ and $m$ are the electron's orbital quantum number and projection number respectively, $n^* = Z \sqrt{U_H/U_{ion}}$ is the effective principle quantum number, and $U_H$ and $U_{ion}$ are the ionization potential of Hydrogen and ions in a principal quantum number state of the material interacting with the laser, respectively. $Z$ is the charge number after ionization.

Let's consider first ionization in hydrogen-type atoms. The ionization probability expression (\ref{eq:multi-level-W}) reduces to
\[
W(s^{-1}) \approx 1.52e^{15}\frac{4^{n^{\star}}U(eV)}{n^{\star}\Gamma(2n^{\star})}\left(20.5\frac{U^{1.5}(eV)}{E(GV/m)}\right) \times
\]
\begin{equation}
\exp\left(-6.83\frac{U^{1.5}(eV)}{E(GV/m)}\right) 
\label{ADK}
\end{equation} 
where $U(eV)$ is the ionization energy of neutral atoms and $n^{\star} = 3.69Z/\sqrt{U(eV)}$ is the effective	principal quantum number.
At every time step, the number of electron-ion pairs to	be generated is calculated	using fractional ionization formula $ 1-exp\left( W(s^{-1})dt\right)$, which is the solution to the following equation:
\begin{equation}
\frac{dn_0}{dt} = -W(t)n_0(t),
\label{rate}
\end{equation} 
where $n_0$ is the neutral atom density and $W(s^{-1})$ is calculated from (\ref{ADK}). Since the characteristic time scales of relevant laser-plasma interactions are short (on the order of 10 picoseconds), the recombination processes are ignored. 

Most PIC codes utilize Monte-Carlo style routine to choose whether to ionize a neutral particle or not by calculating the rate from equation \ref{ADK} and comparing it with a random number in $[0,1]$ \cite{CHEN2013,Smilei}. SPACE, however avoids this expensive routine, and progressively {\itshape{charges}} the region around each grid point by creating number of electron-ion pairs calculated using the rate equation \ref{ADK}. 

\begin{figure}[h!]
	
	\centering
	\includegraphics[width=0.9\linewidth]{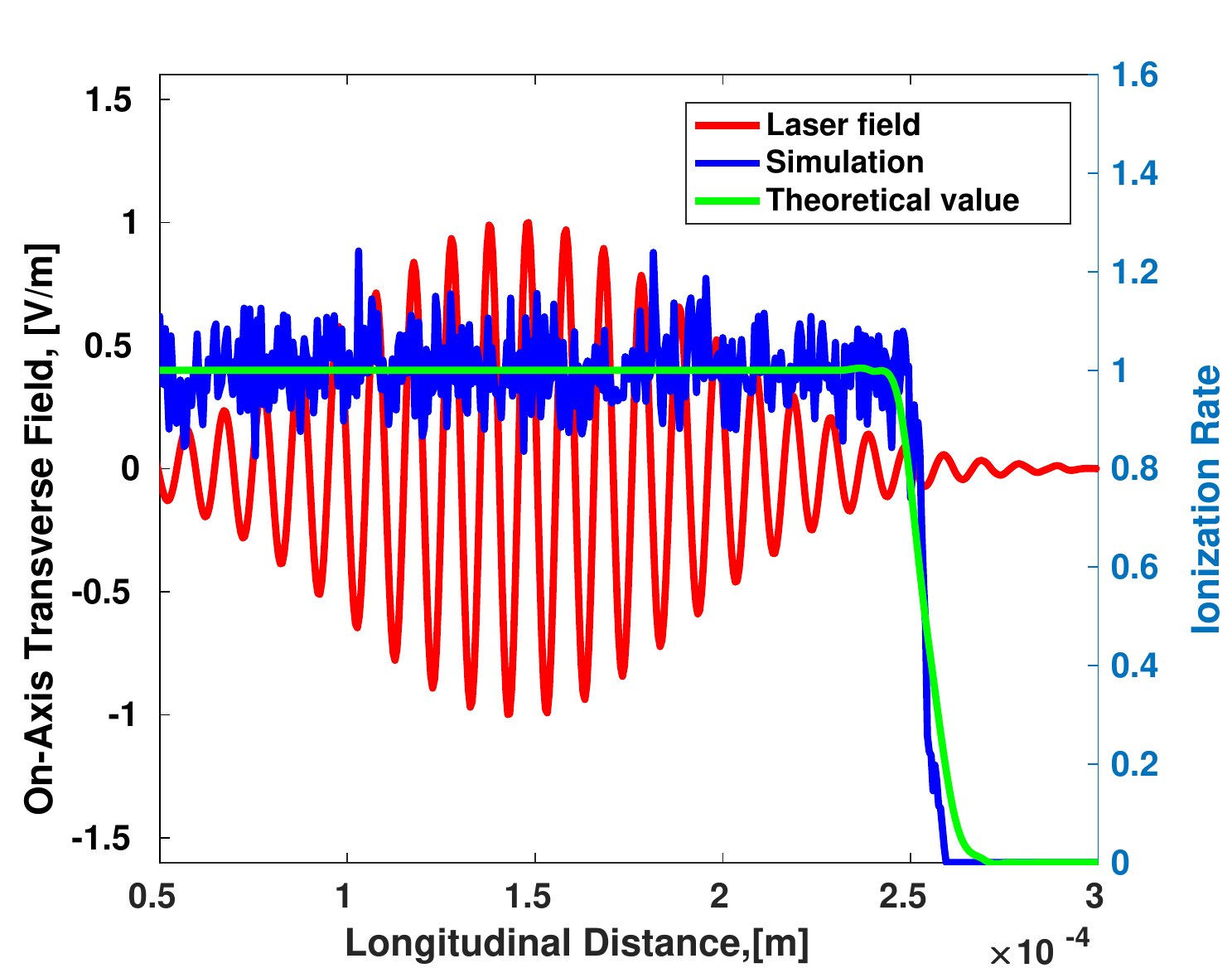}	
	\caption{Ionization of hydrogen by the electric field of CO\textsubscript{2} laser. Red curve shows the normalized electric field of the laser pulse along the direction of propagation. Green curve shows the theoretically predicted rate of ionization. Blue curve shows the normalized plasma distribution along the axis of propagation of the laser pulse as computed by code SPACE. Simulation uses 32 particles per cell and a longitudinal resolution of 20 cells per wavelength of the laser pulse.}
	\label{ionization_ver}
\end{figure} 

Ionization of hydrogen by a CO\textsubscript{2} laser pulse was simulated using SPACE. A linearly polarized laser (wavelength = 10 micrometers) pulse with Gaussian transverse and longitudinal profile was used.  Figure \ref{ionization_ver} shows the comparison between the theoretically predicted rate of ionization and that computed by SPACE. For a longitudinal resolution of 20 cells per wavelength and 32 particles per cell, simulated rate of ionization is in good agreement with the theory.

The following system of equations governs multi-level ionization dynamics in high-atomic-number gases:
\begin{align}
	\frac{dn_0}{dt} &= -W_0 * n_0(t), \nonumber\\
	\frac{dn^+}{dt} &= W_0 * n_0(t) - W_1 * n^+(t), \nonumber\\
	\frac{dn^{(Z-1)+}}{dt} &= W_{Z-2}*n^{(Z-2)+}(t) - W_{Z-1}*n^{(Z-1)+}(t), \\
	\frac{dn^{Z+}}{dt} &= W_{Z-1}*n^{(Z-1)+}. \nonumber
\end{align}
In this system, ionization probabilities are computed using (\ref{eq:multi-level-W}) The system can be written in the matrix form:
\begin{equation}
	\begin{bmatrix} \dot{n_0} \\ \dot{n}^+ \\ \dot{n}^{2+} \\ ... \\ \dot{n}^{(Z-1)+} \\ \dot{n}^{Z+} \end{bmatrix} =
	\label{eq:matrix_form}
\end{equation}
\[	 
	\begin{bmatrix}
		-W_0 & 0    & 0   & ... & 0 & 0 \\
		W_0  & -W_1 & 0   & ... & 0 & 0 \\
		0    & W_1  & -W_2& ... & 0 & 0 \\
		\vdots & ... & ... & \ddots & \vdots & \vdots \\
		\vdots & ... & ... & W_{Z-2} & -W_{Z-1} & 0 \\
		0 & ... & ... & 0 & W_{Z-1} & 0
	\end{bmatrix}
	\begin{bmatrix} n_0 \\ n^+ \\ n^{2+} \\ ... \\ n^{(Z-1)+} \\ n^{Z+} \end{bmatrix}
	\]
We found an analytic solution to this system of equations in terms of the eigenvalues and eigenvectors of the corresponding matrix for constant ionization probabilities \cite{ACheng2021}.
Using the ionization probabilities updated at the beginning of each time step in code SPACE, we compute new analytic solutions in every computational cell and update the population of ionization levels at the end of the time step by the analytic solution. 

The implementation of multiple ionization in SPACE resolves several modeling and computational challenges. The first challenge is associated with strong dependence of ionization probabilities on the orbital quantum numbers and their projections for electrons being ionized. SPACE properly assigns quantum numbers during the ionization process. The second challenge is related to the multiscale nature of ionization.  In particular, the characteristic ionization time scale may be incompatible with the time step of the main code. This problem is effectively resolved by using analytic solutions at each time step. The third challenge is associated with a significant 
increase of memory allocation for computing ionization processes. To store all ionization levels in krypton, 36-floating point numbers must be added to each computational cell (or at least 10-12 numbers, if ionization beyond 10+ is unlikely at given laser parameters). Studying the ionization dynamics of each level under varying electric field strength, we concluded that only 2-3 intermediate levels were populated at any moment of time, with the population of lower and higher levels effectively approaching zero. This allowed us to replace the full system of equations for all ionization levels by a locally reduced-order system containing only 3 or 4 ionization levels. The active ionization levels are shifted to the higher ones independently in each computational cell when the population of the lowest level drops below a prescribed threshold. SPACE code can also use tabulated data base that provides averaged number of electrons based on local values of  the electric field. While this approach is accurate for low-frequency processes, we mostly use analytic solutions for locally-reduced systems of 4-level equations to maintain high accuracy in laser-plasma simulations. Details of the algorithm can be found in \cite{ACheng2021}.

\section{\label{Implement}Implementation}

\subsection{Overview and code structure}
\begin{figure*}[h!]
	\centering=
	\includegraphics[width=\linewidth]{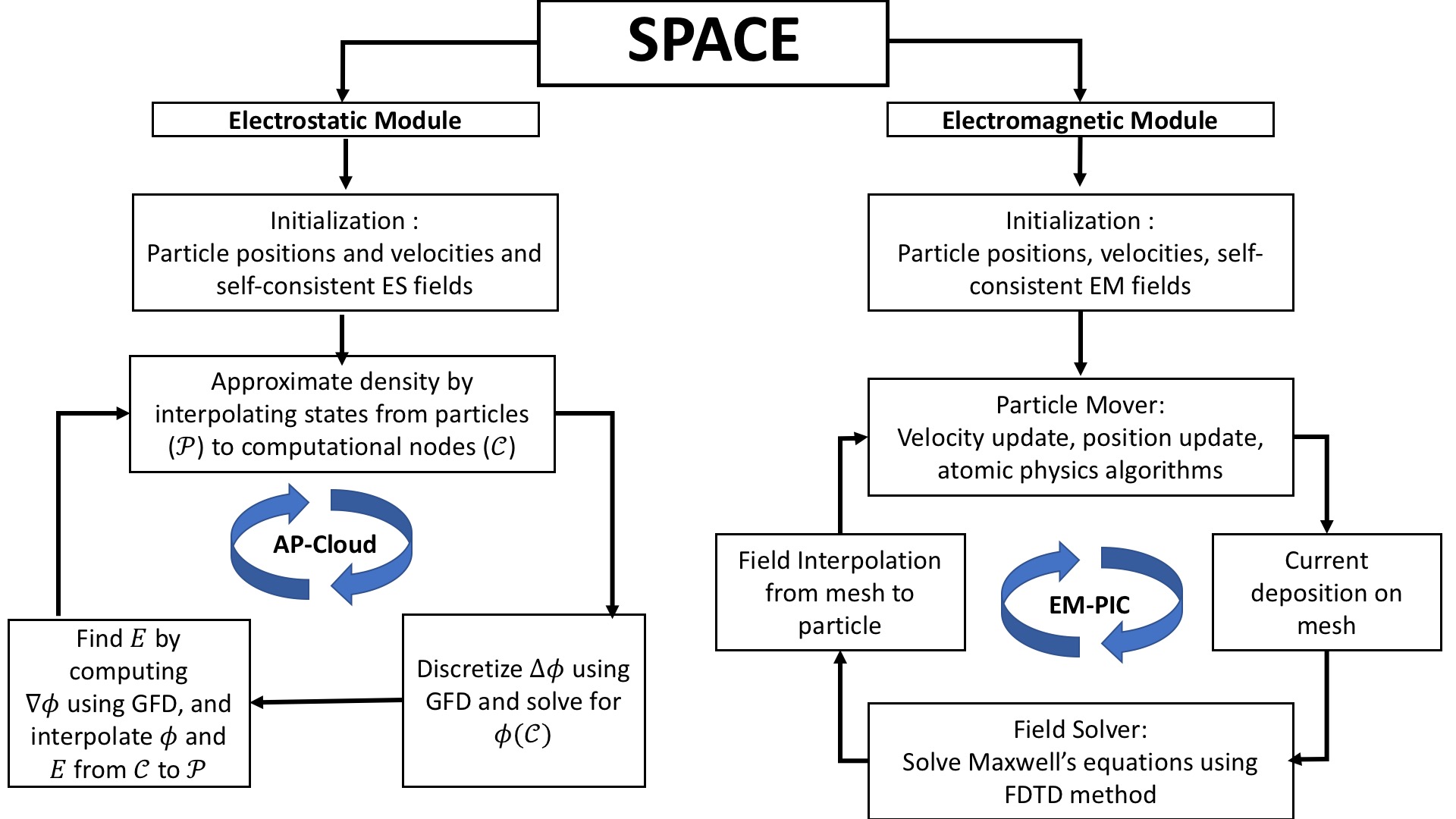}	
	\caption{SPACE code structure}
	\label{structure}	
\end{figure*}

SPACE contains two main modules : the electrostatic (ES) module which implements the AP-Cloud method for solving the Vlasov-Poisson problems and a more traditional electrostatic PIC solver, and the electromagnetic (EM) module which implements the EM-PIC method with atomic physics support. Figure \ref{structure} shows the overall structure of the code. Written in C++, it utilizes an object-oriented design to achieve a flexible and efficient structure.

The ES module consists of two major classes, the \texttt{Interpolator} class which selects computational nodes and interpolates states from computational nodes to the location of macroparticles using the Taylor expansion, and the \texttt{APcloud solver} class which builds linear systems by approximating differential operators in the location of computational nodes using GFD and solves linear systems using fast parallel solver libraries. 

The EM module consists of three major classes: (1) \texttt{FieldSolver} class, (2) \texttt{ParticleMover} class, and (3) \texttt{Controller} class. The \texttt{FieldSolver} class contains the FDTD solver, and the interpolation routines for getting field information at particle locations. The \texttt{ParticleMover} class contains solvers for the Newton-Lorentz equation. This class also includes various physics models describing particle interactions and transformations by atomic physics processes. The code is capable of tracking particles of numerous species. The \texttt{Controller} class controls the above two classes and other miscellaneous classes such as the visualization class, which outputs the electromagnetic fields and particle data in a desired format to use the software VisIt \cite{VisIt} optimized for the parallel remote visualization. Due to the SPACE code structure, implementation of additional physics models and new features is fairly simple. SPACE supports input files written in eXtensible Markup Language (XML).

Both the electromagnetic and electrostatic solvers in SPACE implement a number of boundary conditions. 
In particular, the electromagnetic code includes the perfect electric conductor boundary condition, perfectly matched layer in the longitudinal direction, for simulation of electromagnetic signals passing through the computational domain, and the periodic boundary condition. The AP-Cloud code implements the Dirichlet, Neumann, and periodic boundary conditions for the Poisson problem. It is important to emphasize that AP-Cloud works with computational domains of arbitrary shape.

\subsection{Implementation of atomic physics algorithms}

\begin{figure}[h!]
	\centering
	\includegraphics[width=\linewidth]{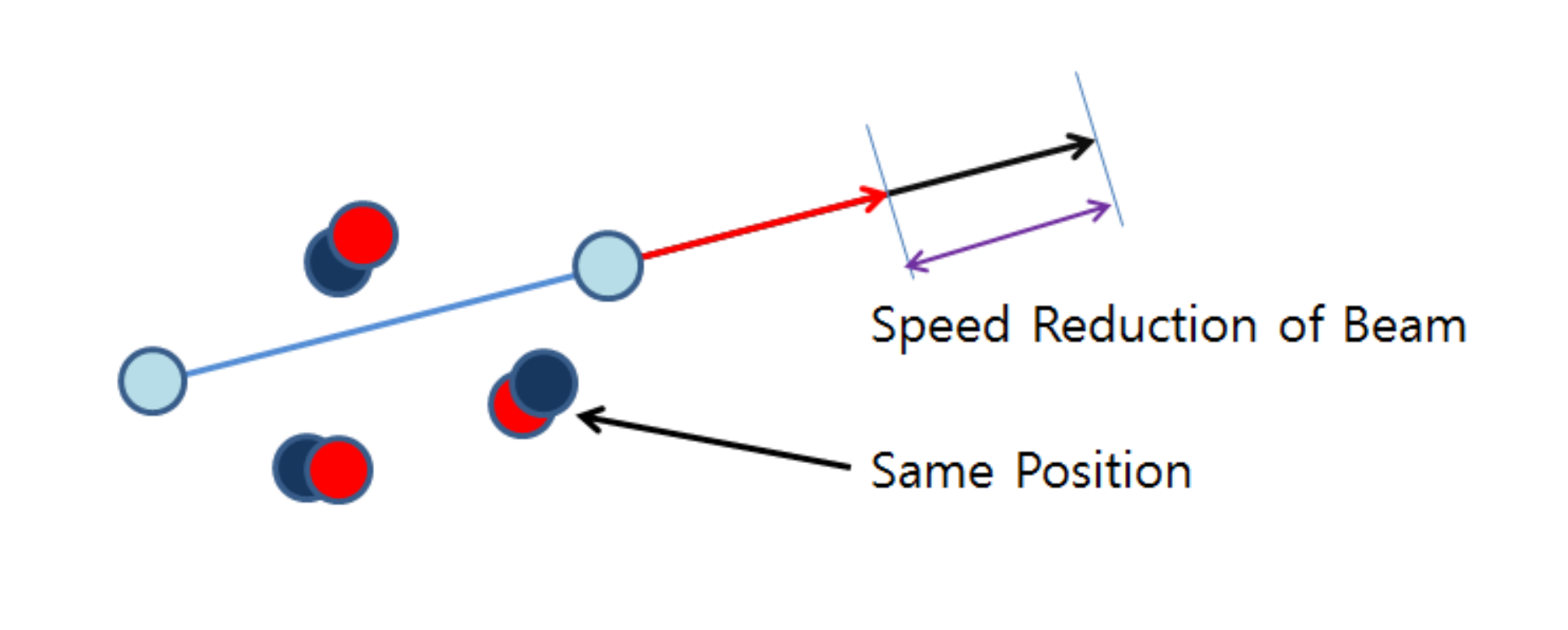}	
	\caption{Schematic of the ionization algorithm by particle beam.}
	\label{ionization_beam}	
\end{figure}

{\bf Generation of plasma macroparticles.}
Two algorithms for the dynamic generation of plasma have been implemented in SPACE. The first algorithm dynamically creates plasma macroparticle pairs and the second algorithm changes
the representing number of macroparticles. Consider an example of neutral gas ionization by a high energy particle beam. As each beam particle passes through the gas, it loses energy and ionizes
the medium by creating electron-ion pairs. The amount of energy lost by the beam through ionization processes, or the beam stopping power, is described by the Bethe-Bloch equation. This process is directly implemented in the
code: the energy loss of every beam macroparticle is computed in real time, and a macro-electron-ion pair is numerically created when the beam particle energy loss exceeds the ionization energy. Each pair of electron and ion
macroparticles is created in the same spatial location to satisfy the initial local charge neutrality. In numerous applications, the mobility of ions is very low throughout the simulation and the motion of ions can be ignored. 
As stationary particles have no effect on the solutions of the Maxwell equations, we need to create only electron macroparticles with zero initial electric field (due to plasma neutrality) in such a case. The initialization of only 
electrons with no electric field is equivalent to the presence of stationary positive ions by the conservative property of the numerical algorithm.  As a result, the electric field due to stationary ions is present in simulations. In simulations with 
strong electromagnetic fields ions are explicitly evolved. The schematic of processes is shown in 
Figure \ref{ionization_beam}. 

\begin{figure}[h!]
	\centering
	\includegraphics[width=0.8\linewidth]{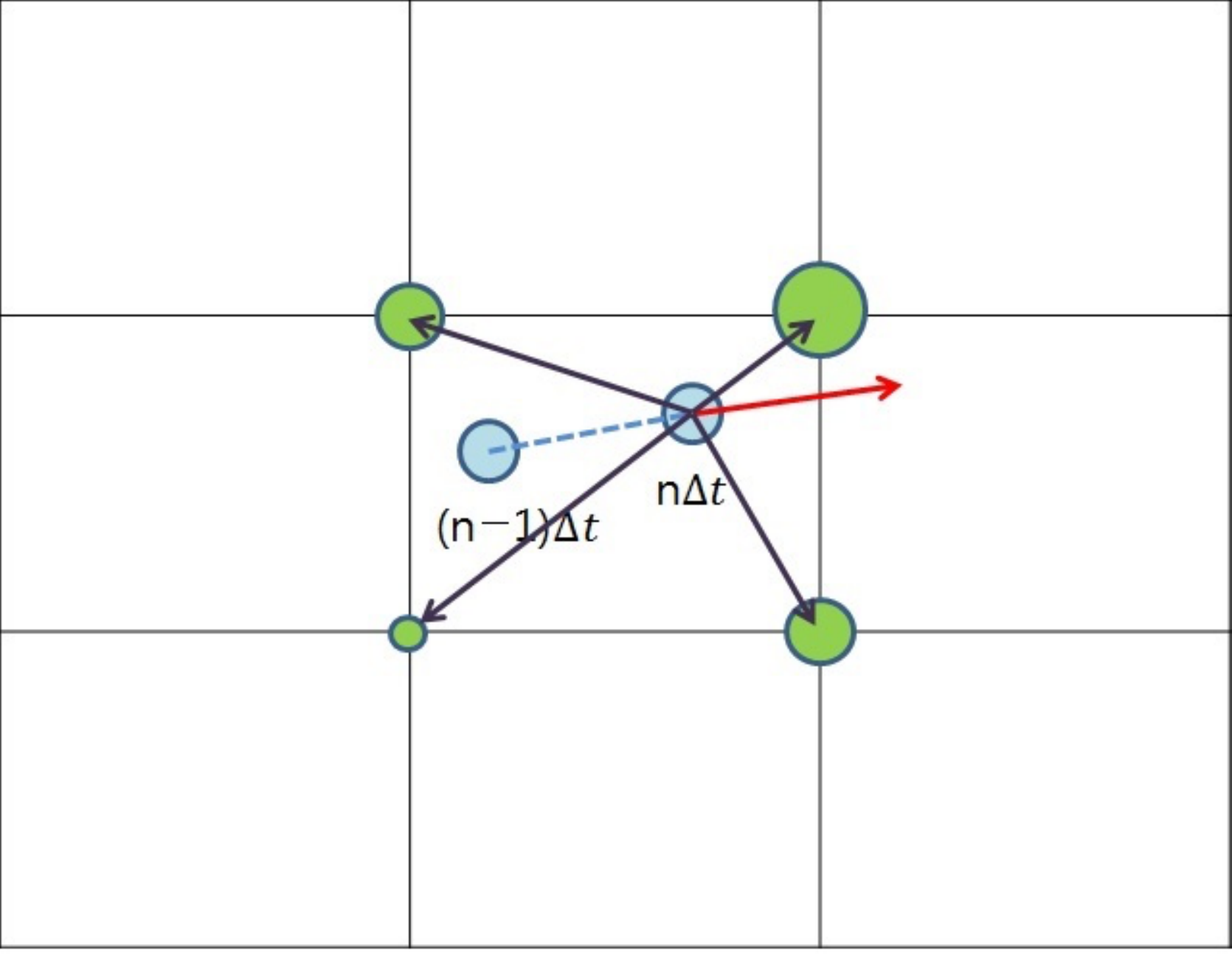}	
	\caption{Schematic diagram illustrating ionization algorithm with variable representing number of plasma macroparticles.}
	\label{ionization_schematic}	
\end{figure}

{\bf Variable representing number.}
In many applications, the plasma density can change by several orders of magnitude during physically relevant times of interest. The method described in the  previous section may lead to numerical difficulties due to the extreme increase of 
the number of macroparticles in high density plasma regions while still achieving poor accuracy in low density regions represented by a small number of macroparticles. This method also leads to numerical oscillations caused by recombination 
processes. Initially overlapping electron and ion macroparticles, created by the ionization process, become spatially separated by dynamic processes at a later time. When a recombination event occurs at some point in space and time, triggered by the computed probability of recombination, the two closest but spatially separated charges must be eliminated, causing numerical noise. This problem is effectively eliminated by using a variable representing number of macroparticles. In this
algorithm, a preset cloud of massless, neutral plasma macroparticles (with zero representing number) is created at the initial simulation time with a number density sufficient for numerical accuracy. At a later time, such
macroparticles are charged proportionally to ionization processes by increasing their representing number, and decreasing the representing number proportional to the recombination rate. The ionization and recombination rates
are computed on the PIC mesh, and the corresponding changes are interpolated to the location of all macroelectrons and are kept on PIC mesh nodes for ion components (as ion particles are not physically present). This algorithm keeps the number of 
computational macroparticles at the optimal level, reduces the numerical noise caused by recombination processes, and eliminates a need for a complex “bookkeeping” algorithm that records the lifetime of every plasma macroparticle and calculates
the probability of its recombination or attachment to a dopant molecule. Figure \ref{ionization_schematic} shows the schematic description of the stopping power computation by a particle beam. By the movement of a beam particle (blue), its energy loss in gas is estimated and 
distributed to the FDTD mesh (green). At the same time, the energy loss of the beam particle is counted and used to update its velocity.  After that, the number of new plasma ionization and recombination events is computed on the 
mesh. Their difference in each mesh point is interpolated to the plasma macroparticles contained within the domain of dependence of the interpolation scheme and used for changing of their representing number.

\subsection{Parallelization of Vlasov-Maxwell Solver}

PIC codes have been used to model complex field-particle interactions which require simulations of large number of particles on a very fine mesh requiring large computational resources. Different parallelization strategies have been used to take advantage of modern, distributed memory machines to enhance the performance. Parallelization of PIC codes is more complicated than purely mesh-based or particle based codes because of the need to account for the interactions between fields on the grid and particles. 

SPACE utilizes the Message Passing Interface (MPI) and OpenMP for parallelization. Two different parallelization methods have been used. In the first method, all the particles, irrespective of their positions, are distributed equally among available MPI processes and each process redundantly keeps the copy of entire mesh. Within each MPI process, particles are equally distributed to available OpenMP threads. The field solver utilizes OpenMP parallelization as well. This method is fairly simple to implement. Particles can be moved independently as each process has access to the entire field information. Particles are distributed uniformly among processes in the beginning and since they never change processes, there is no issue of load balance. The main disadvantage of this method is that it requires a large amount of time for communications between the FieldSolver and the ParticleMover modules. In addition, the maximum number of MPI processes that can be used is limited by memory constraints, limiting scalability.


In the second method, FieldSolver utilizes a spatial domain decomposition and the particles are distributed among the processes based on their positions. Since, the processes now have the field and particle information locally available, the communication time is minimized. Particles can move across subdomain boundaries in this case, and this information is communicated using MPI. Charge conserving  current deposition algorithm requires the field communication near subdomain boundaries, accomplished via buffer cells on either side of each subdomain boundary. Implementation in this case is more complicated compared to the previous method, in particular the implementation of 
current update algorithms near subdomain boundaries.

Because of typical computational domain geometries in laser-plasma interaction problems - long and narrow domains with open boundary conditions in the longitudinal direction, accomplished via the use of the perfectly matched layer algorithm \cite{PML}, a 
one-dimensional domain decomposition with options of uniform and non-uniform divisions among processors has been used in SPACE. In 
simulations with a very non-uniform distribution of particles in the longitudinal direction, 
a finer domain decomposition in denser particle regions leads to a significant performance enhancement.

\begin{figure}[h!]
	\centering
		\includegraphics[width=0.9\linewidth]{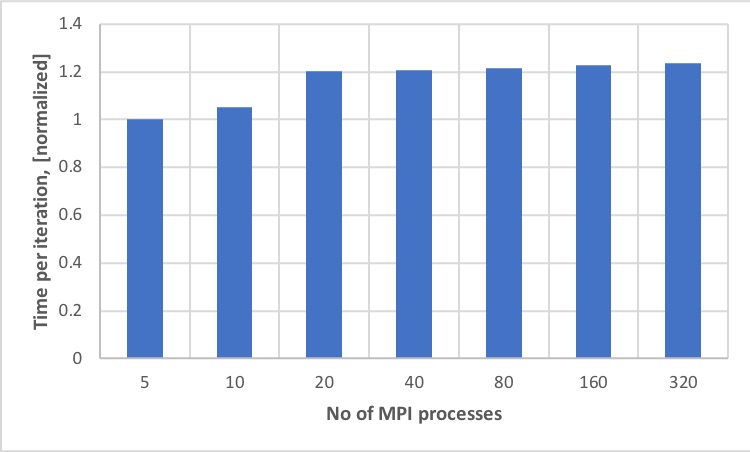}
		\caption{Weak scalability test for 3D laser-plasma interaction simulation.}
		\label{performance}
\end{figure}

We now present  weak scalability  results for performance evaluation of a 3D laser-plasma interaction simulation with a linearly polarized CO\textsubscript{2} laser (energy = 1 J, duration = 2 ps, waist = 20 $\mu m$) interacting with hydrogen plasma with the density of $n_e = 7.5\times 10^{17} cm^{-3}$. Table \ref{table_settings} shows the simulated settings and Figure \ref{performance} shows the average CPU time per iteration in normalized units plotted against the number of MPI cores. An almost perfect scaling was achieved for hundreds of cores.

\begin{table}[h!]
	\begin{center}	
		\scalebox{0.7}{%
		\begin{tabular}{c|c|c|c} 
			\hline
			\textbf{Cells} & \textbf{Particles} & \textbf{MPI processes} & \textbf{Time per iteration} \\
			\hline
			$200 \times 200 \times 200 \times 2^0$ & 80 million & 5 & 1.00 \\
			$200 \times 200 \times 200 \times 2^1$ & 160 million & 10 & 1.05\\
			$200 \times 200 \times 200 \times 2^2$ & 320 million & 20 & 1.20\\
			$200 \times 200 \times 200 \times 2^3$ & 640 million & 40 & 1.21\\
			$200 \times 200 \times 200 \times 2^4$ & 1.28 billion & 80 & 1.21\\
			$200 \times 200 \times 200 \times 2^5$ & 2.56 billion & 160 & 1.23\\
			$200 \times 200 \times 200 \times 2^6$ & 5.12 billion & 320 & 1.23\\
			\hline
		\end{tabular}}
		\caption{Scalability test settings.}
		\label{table_settings}
	\end{center}
\end{table}

\subsection{Parallelization of Vlasov-Poisson Solver}

The AP-Cloud code uses K-tree data structures to store particles and selects centers of K-tree cells as computational nodes. Major K-tree algorithms, such as constructing, refining, and searching, significantly affect performance of the code. The AP-Cloud code utilizes p4est ("parallel forest of K-trees")\cite{BursteddeWilcoxGhattas11}, a parallel library that implements a dynamic management of a collection of adaptive K-trees on distributed memory supercomputers. Given a distribution of macroparticles, the \texttt{Interpolator} calls p4est parallel routines to construct the K-tree, adaptively refine the K-tree until the error balance criterion (\ref{errorbalance}) is met by all K-tree cells, and enforce the 2:1 K-tree balance to improve the smoothness in the placement of computational particles. In order to achieve computational load-balance, K-tree cells are distributed evenly among processes, together with macroparticles inside them. 
This approach optimizes the main computational task: discretization of differential operators and solving of linear systems for computing stencil coefficients in the location of computational nodes.
We use the p4est "Ghost" routine to collect several layers of ghost cells (off-process K-tree cells touching the process boundary) to give the complete parallel neighbourhood information for processes. Hence, the neighbour searching algorithm involved in building GFD stencils is performed locally, eliminating the difficulties of point-to-point transfers of numerical information.

The \texttt{APcloud solver} class uses PETSc \cite{petsc-efficient} library of Krylov subspace solvers and preconditioners to solve the large sparse linear systems resulting from the AP-Cloud discretization of a Vlasov-Poisson problem. The above mentioned partitioning also results in the optimal matrix storage among processes for PETSc matrix initialization routines, since each computational node corresponds to a row in the matrix. The interpolation of states from nodes to macroparticles is also performed locally because K-tree cells and macroparticles inside them are stored on the same processes.

\begin{figure}[h!]
	\begin{subfigure}{\linewidth}
		\centering
		\includegraphics[width=\linewidth]{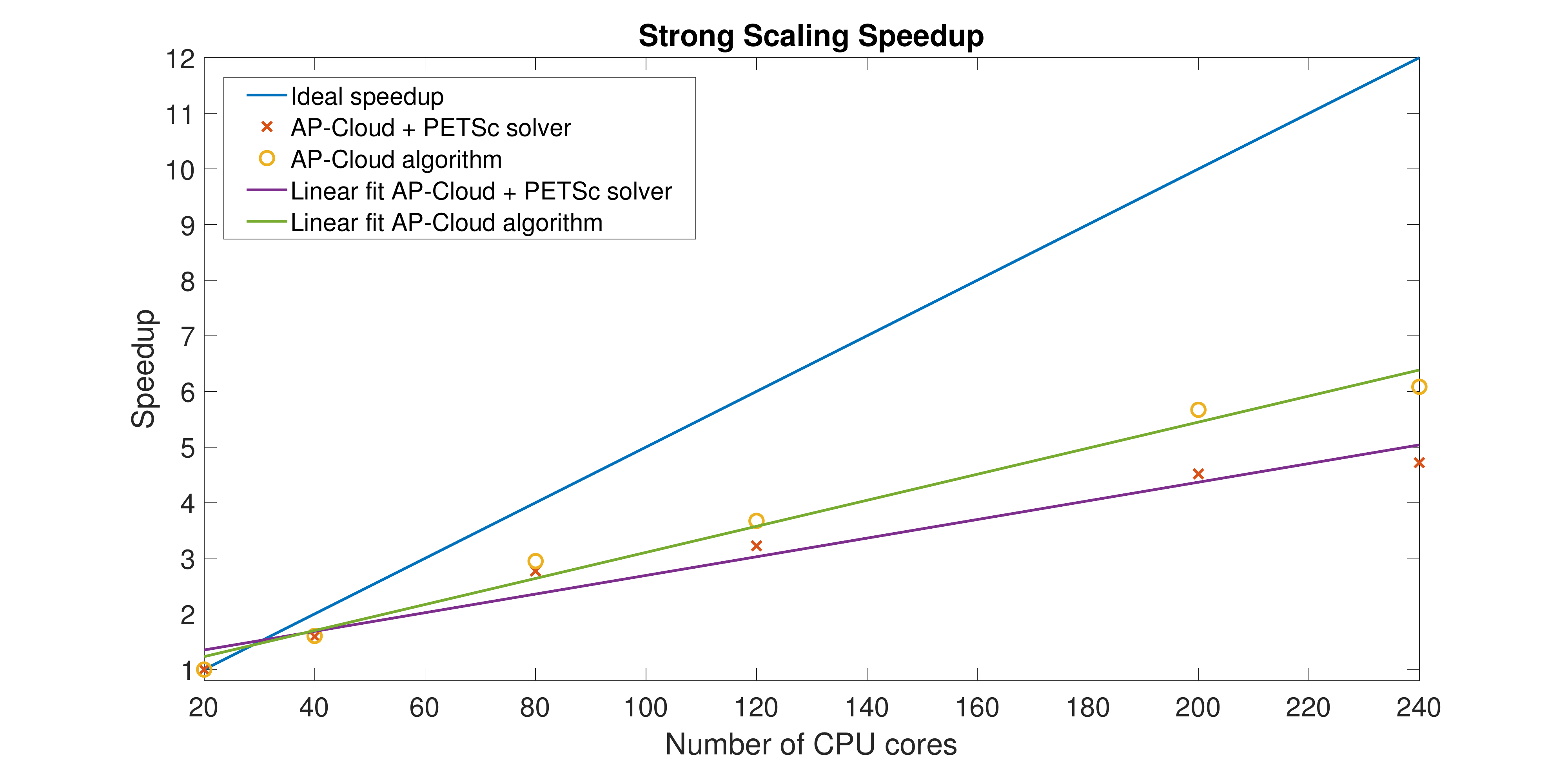}
		\caption{AP-Cloud strong scaling}
		\label{apcloud_strong_scaling}
	\end{subfigure} 
	\begin{subfigure}{\linewidth}
		\centering
		\includegraphics[width=\linewidth]{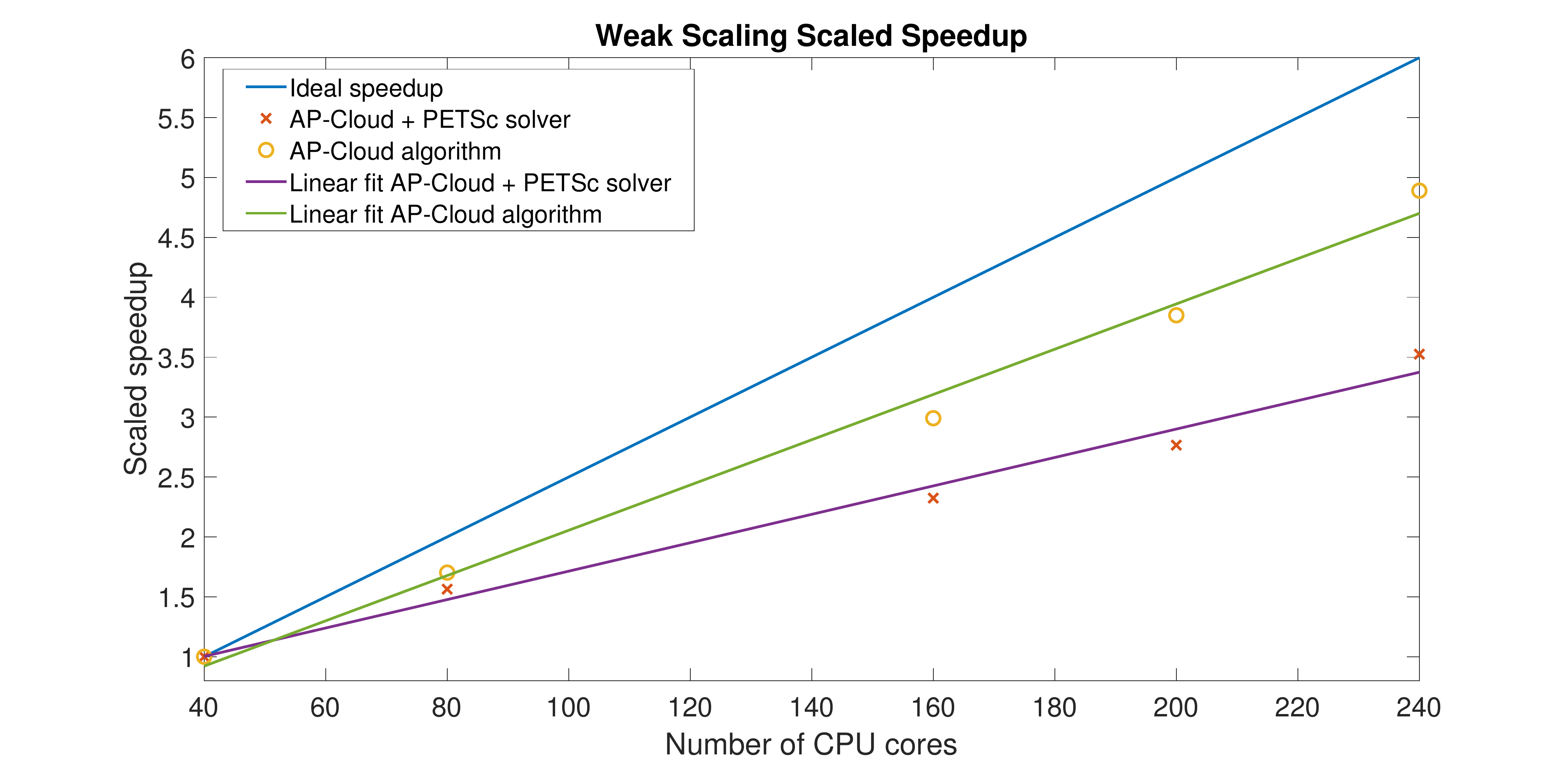}
		\caption{AP-Cloud weak scaling}
		\label{apcloud_weak_scaling}
	\end{subfigure} 
	\caption{Speedup of AP-Cloud code using a Poisson problem with the Gaussian particle distribution. For strong scaling studies, $10^8$ macroparticles and 5,809,343 computational nodes were used.}
	\label{apcloud_scaling}
\end{figure}

We first investigate the strong scalability of AP-Cloud code. We evaluated the performance of simulation for the following Poisson problem: a Gaussian distribution of $10^8$ macroparticles (mean $\mu=[0,0,0]$ and standard deviation $\sigma=0.01$) were generated in a unit cubic domain with Dirichlet boundary condition along all boundaries. The distrubuiton of physical macroparticles resuted in 5,809,343 computational nodes. The scalability of just AP-Cloud algorithms (without the final PETSc solver step) were tested as well as the scalability of the complete time step including the PETSc solver and the results are plotted in Figure \ref{apcloud_strong_scaling}.
While the slope of the  linear fit to numerical data points differs from the ideal speedup, the numerical results demonstrate a liner speedup, without reaching saturation at 240 cores.  For the weak scaling, we performed the same test problem on 240 CPU cores and scaled down both the number of macroparticles and computational nodes proportionally to the reduction of CPU cores. Figure \ref{apcloud_weak_scaling} shows the corresponding numerical results for the
weak scaling study. As expected, the weak scaling is closer to the ideal scaling and it also does not reach the saturation at 240 cores.

\section{\label{Apply} Representative Applications}

SPACE is currently being used in a variety of applications in the area of particle accelerator design as well as fundamental and applied problems of laser-plasma interaction. 

The AP-Cloud code is extensively used at the Brookhaven National Laboratory for the study of coherent electron cooling cooling (CeC), a novel technique developed for
rapid cooling of high-energy, high-intensity hadron beams \cite{LitvinenkoDerbenev2009}.
CeC consists of three main components: a modulator,
where each ion imprints a density wake on the electron
distribution, a free electron laser (FEL) or another electromagnetic device as an amplifier,
where the density wakes are amplified, and a kicker, where
the amplified wakes interact with ions, resulting in
dynamical friction for the ion that leads to cooling of ion beams. Comprehensive numerical studies of the modulator, performed in \cite{Ma2018}, include various verification tests and their comparison with simplified analytic solutions as well as detailed simulations at conditions of the BNL experiment. Simulations 
of coherent electron cooling with two types of amplifiers, the free electron laser and plasma cascade amplifier, were performed in
\cite{Ma2019}.  In both cases, AP-Cloud was used as a component of a coupled simulation. To evaluate the performance of the 
plasma cascade amplifier (PCA), AP-Cloud simulation of the modulator and the kicker was coupled to the electromgnetic SPACE simulation of PCA. AP-Cloud was also coupled to the GENESIS free electron laser code for a start-to-end simulation of CeC with the FEL amplifier.

The capability of SPACE to resolve long-time-scale atomic physics transformaitons in gases interacting with high energy particle beams was critical for the simulation support of the experimental program on the high-pressure hydrogen gas-filled RF cavity (HPRF cavity) in the Mucool Test Area (MTA) at Fermilab. The project studied processes relevant to muon cooling devices. 
We have investigated the plasma dynamics in the RF cavity including the process of power dump by plasma (plasma loading), recombination of plasma, and plasma interaction with dopant material. By comparison with experiments in the MTA, simulations 
suggest several unknown properties of plasma such as the effective recombination rate, the electron attachment time on dopant molecule, and the ion – ion recombination rate in the plasma \cite{Yu2017}. As muon beams were not
available at Fermilab, all experiments used proton beams. SPACE simulations, validated agains experimental data from the HPRF cavity driven by proton beams, performed prediction for muon beams \cite{Yu2018}.

SPACE has been extensively used for the study of laser-matter interactions and wakefield acceleration.
Numerical studies of the interaction of a CO\textsubscript{2} laser with hydrogen
jets have been performed in \cite{Kumar2019} as a part of laser wakefield acceleration program
at BNL’s Accelerator Test Facility. The upgraded laser system is
capable of delivering pulses with parameters suitable for self-modulated laser wakefield acceleration (SM-LWFA). Simulations reproduced both Stokes and anti-Stokes shifts in the spectrum of the pump laser, similar to those observed in
experiments in the spectrum of the probe laser. Good agreement has been achieved with the experiments on the effect of variation in gas density on Stokes/anti-Stokes intensity. In addition, self-injection and trapping of electrons into the self-modulated wakes have been observed and analyzed.
In our recent work \cite{Kumar2021}, long wavelength infrared laser-driven plasma wakefield accelerators were investigated in the self-modulated laser wakefield acceleration  and blowout regimes via SPACE simulations. Simulation results showed that in the SM-LWFA regime, self-injection
arises with wave breaking, whereas in the blowout regime, self-injection is not observed under the simulation conditions. The wave
breaking process in the SM-LWFA regime occurs at a field strength that is significantly below the 1D wave-breaking threshold. This process intensifies at higher laser power and plasma density and is suppressed at low plasma densities ($\le 1\times 10^{17}$ cm$^{-3}$). The produced electrons show spatial modulations with a period matching that of the laser wavelength, which is a clear signature of direct laser acceleration. SPACE has also been used for simulations of laser-plasma wakefields at realistic supersonic hydrogen jet conditions \cite{Kumar-thesis} using input data from the Lagrangian particle hydrodynamic code \cite{Samulyak2018}.

\section{Conclusion}
SPACE, a parallel, relativistic, three-dimensional particle-in-cell code for the simulation of electromagnetic fields, relativistic particle
beams, and plasmas is described in this paper.

The electromagnetic module of SPACE implements a parallel, fully relativistic, second-order FDTD algorithm for electromagnetic fields and particle beams.  In addition to the standard PIC algorithm, SPACE includes an efficient method for highly relativistic
beams in nonrelativistic plasma, support for simulations in relativistic moving
frames, and a special data transfer algorithm for transformations from
moving frames to laboratory frames that resolves the problem of individual times
of particles.

SPACE includes efficient novel algorithms to resolve
atomic physics processes such as multi-level ionization of plasma atoms, recombination,
and electron attachment to dopants in dense neutral gases. The multi-level ioinization algorithm in high-atomic-number gases is highly efficient, featuring several improvements compared to the previously published methods. In particular, it effectively resolves the multiscale nature of ionization processes, which may occur at time scales significantly different compared to the main code time step of the code, by using analytic solutions to the system of differential equations describing ionization evolution. 
The code eliminates the need to store multiple ionization states in every cell of the numerical mesh by using a locally-reduced system of equations. 

For the electrostatic (Vlasov-Poisson) problems, SPACE implements a highly adaptive particle-based method, called Adaptive
Particle-in-Cloud (AP-Cloud) in addition to the more traditional PIC algorithm.
AP-Cloud eliminates the traditional Cartesian mesh of PIC and replaces it with an
adaptive octree data structure. It adaptively selects computaitonal nodes in a cloud of physical macroparticles in such a way that the discretization error of the differential operator is of the same magnitude as the Monte-Carlo error of the source evaluation, thus solving the problem with an optimal numerical resolution. It is also free of the artifacts typical for the (unmitigated) adaptive mesh refinement in an electrostatic PIC method. AP-Cloud is applicable to arbitrary geometrical shape domains and it is parallelized for distributed memory supercomputers using p4est, a parallel library that implements a dynamic management of a collection of adaptive K-trees.

SPACE has been extensively used in a variety of applications in the area of plasma physics and accelerator design, in particular for simulations of coherent electron cooling of relativistic ion beams, interaction of proton beams with high density gas in support of the muon cooling experimental program, 
and the laser-plasma wakefield acceleration.

In the future, we plan to enhance SPACE with models and algorithms for high-density collisional plasma
and extend the spectrum of its applications.


\vskip10mm

{\bf Acknowledgments}

This work was supported by Grant No. DE-SC0014043 funded
by the U.S. Department of Energy, Office of Science, High Energy
Physics.


\bibliography{SPACE_CPC}

\end{document}